\documentclass[conference]{IEEEtran}
\IEEEoverridecommandlockouts
\usepackage{cite}
\usepackage{amsmath,amssymb,amsfonts}
\usepackage{algorithmic}
\usepackage{graphicx}
\usepackage{textcomp}
\usepackage{xcolor}
\def\BibTeX{{\rm B\kern-.05em{\sc i\kern-.025em b}\kern-.08em
    T\kern-.1667em\lower.7ex\hbox{E}\kern-.125emX}}
\begin{document}

\title{Classification of Common Waveforms Including a Watchdog for Unknown Signals\\
\thanks{We wish to acknowledge the US Army Research Laboratory for their funding of this project.}
}

\author{\IEEEauthorblockN{C. Tanner Fredieu$^1$, Justin Bui$^2$, Anthony Martone$^3$, Robert J. Marks II$^2$, Charles Baylis$^2$, R. Michael Buehrer$^1$}
\IEEEauthorblockA{\textit{$^1$Wireless@VT, Bradley Department of Electrical and Computer Engineering, Virginia Tech, Blacksburg, VA, USA } \\
\textit{$^2$Department of Electrical and Computer Engineering, Baylor University, Waco, TX, USA}\\
\textit{$^3$US Army Research Laboratory, Adelphi, MD, USA}
\\
Email: christianf@vt.edu, justin\_bui@baylor.edu, anthony.f.martone.civ@mail.mil, \\robert\_marks@baylor.edu, charles\_baylis@baylor.edu, rbuehrer@vt.edu}

}

\maketitle

\begin{abstract}
In this paper, we examine the use of a deep multi-layer perceptron model architecture to classify received signal samples as coming from one of four common waveforms, Single Carrier (SC), Single-Carrier Frequency Division Multiple Access (SC-FDMA), Orthogonal Frequency Division Multiplexing (OFDM), and Linear Frequency Modulation (LFM), used in communication and radar networks. Synchronization of the signals is not needed as we assume there is an unknown and uncompensated time and frequency offset. An autoencoder with a deep CNN architecture is also examined to create a new fifth classification category of an unknown waveform type. This is accomplished by calculating a minimum and maximum threshold values from the root mean square error (RMSE) of the radar and communication waveforms. The classifier and autoencoder work together to monitor a spectrum area to identify the common waveforms inside the area of operation along with detecting unknown waveforms. Results from testing showed the classifier had 100\% classification rate above 0 dB with accuracy of 83.2\% and 94.7\% at -10 dB and -5 dB, respectively, with signal impairments present. Results for the anomaly detector showed 85.3\% accuracy at 0 dB with 100\% at SNR greater than 0 dB with signal impairments present when using a high-value Fast Fourier Transform (FFT) size. Accurate detection rates decline as additional noise is introduced to the signals, with 78.1\% at -5 dB and 56.5\% at -10 dB. However, these low rates seen can be potentially mitigated by using even higher FFT sizes also shown in our results.
\end{abstract}


\section{Introduction}
Given the ever-increasing demand for wireless spectrum (especially below 6 GHz), spectrum sharing between communications and radar systems has become an important research topic. Although this allows communication networks to operate in traditional radar bands, it does require advanced methods for enhanced monitoring and management of spectrum to limit new sources of interference. Many of the current solutions include the use of dynamic spectrum access (DSA) [8] and/or cognitive radio [8], [9]. Many approaches and research efforts [5], [8], [10] are already underway to solve the challenges of these techniques such as implementing spectrum sensing, sharing, and management. Most of these solutions have involved the use of machine learning techniques [1], [3], [4], [5], [6], [7], [8] due to their classification abilities. 

Neural networks, especially deep networks have become the preferred method for classifying different components or entire signals in the wireless community, as well-documented in [1], [3], [6], [7]. The most popular neural network architecture for the classification of communication and radar signals is the convolutional neural network (CNN) [1], [3], [6], [7]. This architecture has demonstrated good performance in signal and modulation identification especially in wireless standards classification [3], [7]. The most recent work with waveform and modulation classification using CNNs is detailed in [1]. Kong, Jung, and Koivunen use two CNNs to classify generalized waveform types and modulations with additive white Gaussian noise (AWGN) and multi-path fading present using a Fourier synchrosqueezing transform (FSST) with size 1024 as the input. The classified waveforms included SC-radar, SC, and multi-carrier or OFDM. These models have demonstrated accuracy rates of 100\% for the waveforms with AWGN present and $>$95\% with fading present over a 0 dB to 20 dB SNR range. Although, CNNs have been proven time and time again, they are not the only method that can be used. Deep feed-forward and Long Short Term Memory (LSTM) networks have also shown much promise in certain areas [5]. The primary disadvantage of CNNs and LSTMs is the training requirements of these networks. 

In this paper, we will examine a different technique using an effective, less-complex deep feed-forward model to achieve better results for the classification of four common waveforms used in radar and communication networks. Signals will contain more signal impairments comprised of AWGN, multi-path fading, frequency offset, phase offset, and IQ imbalance. We also propose the concept of using another model to reject signals unknown to the classifier using an convolutional autoencoder watchdog.

Autoencoders have gained traction recently in their ability to better recover signals in noise as well as detect possible anomalies [12], [13]. Autoencoders are a type of neural network that attempts to copy the original data from the input to the output of the network. This is achieved by the hidden layers learning an encoding of the input data [11]. The autoencoder is made up of an encoder that learns a low-dimensionality feature encoding from the input data, as stated previously, and a decoder that takes the encoding and attempts to reconstruct the original data. The error between the reconstructed data and the original data is typically characterized using the RMSE. For the purpose of anomaly detection, the RMSE of the output value for the signal is compared against a pre-determined threshold found from the training of the network. If the value of the signal lies within the threshold region, then the signal is considered to be known. If outside the region, the signal is considered to be unknown. Autoencoder models have been used for this purpose in detecting outlier images such as with MNIST datasets [12]. This approach has the potential for detecting unknown signals within the spectrum environments as will be illustrated in this paper.

\section{Neural Network Architectures}
The proposed system for spectrum monitoring consists of two neural network architectures. A deep feed-forward network with 4 output nodes to classify the four waveforms, and a deep convolutional autoencoder that uses RMSE thresholds to detect unknown waveforms within the environment. The autoencoder will receive the signals first and only signals identified as known will be sent to the classifier for classification. 

\subsection{Optimization and Regularization}\label{AA}
Regularization, a method to prevent machine learning models from over-fitting on data during training [14], is performed using different techniques for each model. Dropout is used for the classifier as it proves to be the most effective during optimization. This technique causes certain nodes at each layer to drop randomly from use to reduce the co-adaptions formed from backpropagation during training [15]. 2$^n$ number of possible reduced neural networks are formed sharing the weights where n is the number of nodes, and an approximate averaging method is used to simplify all the trained networks into a single one [15].

L2 regularization was found to be optimal with the anomaly detector. L2 regularization, also referred to as weight decay, uses a penalty term added to the cost function to prevent the values of the network weights from becoming too large [14]. The value of dropout, value of L2, number of nodes, number of layers, and learning rate were all determined after going through multiple optimizations to find the ideal parameters.

The optimizer is used to apply the gradient to the network allowing the network to learn. Adam, Adamax, and SGD optimizers were compared with Adamax showing the highest accuracy performance for both models. Adam optimization, as well as the general concept behind Adamax version of Adam, is documented in [2] which introduced the technique.

\subsection{Deep Feed-Forward (Classifier)}\label{AA}
As stated previously, the waveform classifier is created by using a deep feed-forward network architecture outlined in Figure 1. The specific network architecture consists an input using the magnitude of the 4096-FFT of each signal, dropout layers added for regularization to prevent over-fitting, and five layers including four hidden layers consisting of 64, 100, 32, 16, and 4 nodes, respectively. Network optimization is performed with an Adamax using default settings. Binary cross-entropy is used for the loss function. Rectified linear unit (ReLu) activation functions are used for each hidden layer while sigmoid activation is used in the output layer to perform classification.

\begin{figure}[t]
\centerline{\includegraphics[scale=0.335]{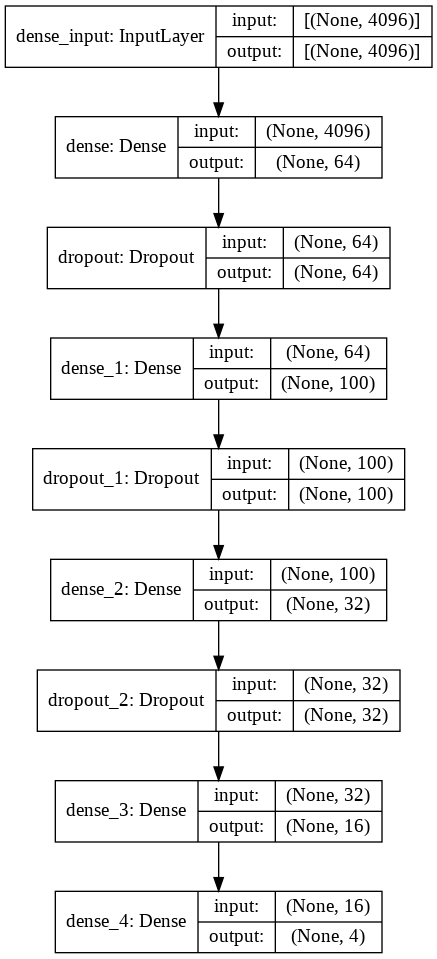}}
\caption{Waveform classifier architecture}
\label{fig1}
\end{figure}

\subsection{Convolutional Autoencoder (Watchdog)}
The general architecture structure of an autoencoder model is illustrated in Figure 2. The power spectral density (PSD) converted to dB of each signal is used as the input for the network. The input size depends on the FFT size. The network is comprised of two parts consisting of the encoder, which learns the representation of the signal through unsupervised learning, and the decoder, which takes the representation to recreate the signal [11]. The layout is symmetrical as the encoder uses three convolutional layers connected to a fully-connected dense layer while the decoder uses a fully-connected dense layer connected to four convolutional transpose layers. The numbers filters in each convolutional layer of the encoder are 256, 128, 32, and 16 for the number of nodes in the fully-connected layer, respectively. The decoder nodes are the inverse of the encoder order with the output layer being a convolutional transpose layer with 1 filter. For the regularization, the encoder and decoder both use L2 regularization at the convolutional and convolutional transpose layers to prevent over-fitting. This was set to 0.1. Network optimization is also performed using an Adamax optimizer with learning rate set to 0.1, and binary cross-entropy used as loss function. ReLu is also used as the activation for each layer except the output where sigmoid activation is used.

\begin{figure}[t]
\centerline{\includegraphics[scale=0.47]{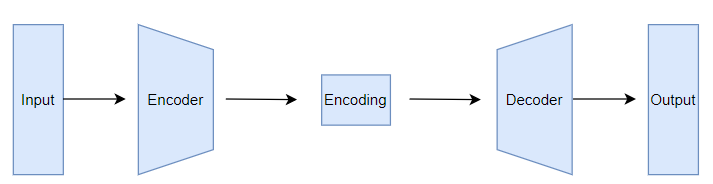}}
\caption{Block diagram of autoencoder}
\label{fig2}
\end{figure}

\section{Training and Data}

In this section, we discuss how the models are trained along with a description of the training and testing datasets used.  The training and evaluation datasets have been normalized between 0 and 1 as with other methods of machine learning due to the models being able to handle this range more efficiently. All training samples are 50 MHz bandwidth signals with a 100 MHz sampling rate lasting 1 ms.

\subsection{Training Dataset}
The training data composed of the four waveform classes were generated using MATLAB scripts inside the MATLAB R2018b program. The training dataset is used for training both the classifier and anomaly detector containing 40,000 signals of IQ data with each of the four waveform classes containing 10,000 signals. For the communication waveforms, each signal class is then broken down further into eight modulation types. These included BPSK, QPSK, 16-PSK, 64-PSK, 4-QAM, 16-QAM, 64-QAM, and 256-QAM. The distribution of among the dataset is illustrated in Table 1.

Signal impairments are added to each waveform class to provide more real-world environmental effects. These impairments include AWGN ranging from -20 dB to 20 dB, carrier phase offset ranging from -$\pi$ to $\pi$, frequency offset ranging from -5000 Hz to 5000 Hz based on the sampling rate of 100 MHz used for all samples, IQ imbalance ranging from 0 dB to 3 dB, and Rayleigh and Rician Fading. The training data used for the anomaly detector contained all the aforementioned impairments, but the AWGN impairments were ranged from 0 dB to 20 dB. This was because it was found during training that the RMSE threshold would not converge to a common threshold if it was trained on signals containing AWGN below 0 dB. 

Training data was preprocessed for each model. The magnitude of the 4096-FFT of each signal was taken for input into the classifier. The anomaly detector used the PSD of a signal converted to dB with FFT sizes of 4096, 8192, and 16384 over the whole dataset to examine the impact of feature size on anomaly detection performance. This same pre-processing is also performed on the test datasets discussed in the next section.

\begin{table}[t]
\caption{Number of training signals with modulation type in each communication waveform class}
\begin{center}
\begin{tabular}{|c|c|c|c|}
\hline
\cline{2-4} 
\textbf{Modulation} & \textbf{\textit{SC}}& \textbf{\textit{SCFDMA}}& \textbf{\textit{OFDM}} \\
\hline
BPSK & 1250 & 1250 & 1250 \\
\hline
QPSK & 1250 & 1250 & 1250 \\
\hline
16-PSK & 1250 & 1250 & 1250 \\
\hline
64-PSK & 1250 & 1250 & 1250 \\
\hline
4-QAM & 1250 & 1250 & 1250 \\
\hline
16-QAM & 1250 & 1250 & 1250 \\
\hline
64-QAM & 1250 & 1250 & 1250 \\
\hline
256-QAM & 1250 & 1250 & 1250 \\
\hline
\end{tabular}
\label{tab1}
\end{center}
\end{table}

\subsection{Testing Datasets}
The testing datasets contain a total of 3200 signals each composed of the SC, SCFDMA, OFDM, and LFM signals along with FM, AM, Bluetooth low energy (BLE) 5.0, and white noise signals not included in the training dataset. Each waveform class contains 400 test signals. The test dataset used for the classifier contained all the aforementioned signal impairments. Two test datasets were used for the anomaly detector. The first contained only AWGN impairments in the range -10 dB to 20 dB while the second contained all the signal impairments using the same AWGN range as the first. AM and FM signals are 50 MHz bandwidth signals with a 100 MHz sampling rate lasting 1 ms. Bluetooth signals are 2 MHz bandwidth signals with a 125 MHz sampling rate lasting 10 ms produced by one or more Bluetooth devices. Each signal also contains frequency hopping throughout the signal that can potentially show a higher bandwidth than 2 MHz. White noise signals are sampled the same as training data and last for 1 ms.

\subsection{Training Methods}
Each model was trained using the same values for epoch and batch size. These values were set to 50 for epoch and 128 for the batch size. The dropout rate was set at 0.2 in all dropout layers shown in Figure 1 to provide regularization. No weight decay or other regularization methods were used. 

For the anomaly detector, dropout regularization was not used. However, L2 regularization set to 0.1 was used at each convolutional and convolutional transpose layer on the encoder and decoder. Other regularization methods were not used.

\section{Results}
In this section, we discuss the results of the system performances classifying various signals belonging to SC, SC-FDMA, OFDM, and LFM along waveform types not previously seen by the classifier.

\subsection{Waveform Classifier}
The overall performance of the classifier tested against signals with all impairments is illustrated in Figure 3. Due to the classifier's ability to correctly identify LFM at 100\% accuracy even at low SNR values, Figure 3 shows the classification performance across SNR values of each individual waveform type so the full scope of the classifier's abilities can be illustrated. The classifier is able to achieve 100\% across all waveforms beginning at 0 dB and achieve above 85\% accuracy for all waveforms at -5 dB. Most of the error results from classifications of OFDM and SC-FDMA as shown in the confusion matrix illustrated in Figure 7.

Figure 4, Figure 5, and Figure 6 are shown to illustrate the minimal effect the signal impairments have on the classifier's performance. This is also illustrated by the comparison of the confusion matrices of the classifier in Figure 7 and Figure 8 with all impairments present and signals with only AWGN, respectively. Phase offsets have almost no effect on classification while frequency offsets have approximately less than 5\% accuracy difference in the lowest SNR value and less than 2\% in other cases. IQ imbalance has no effect on classification rates.

\begin{figure}[t]
\centerline{\includegraphics[scale=0.4]{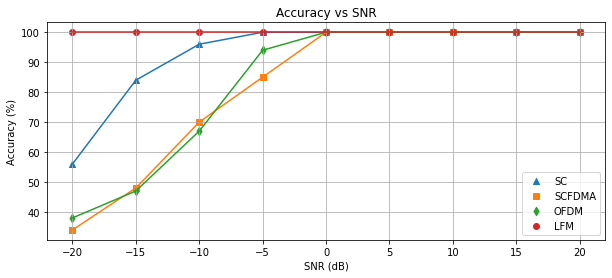}}
\caption{Classifier performance}
\label{fig3}
\end{figure}

\begin{figure}[t]
\centerline{\includegraphics[scale=0.32]{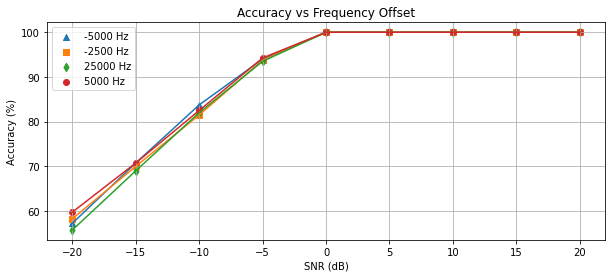}}
\caption{Impact of frequency offset on classifier performance}
\label{fig4}
\end{figure}

\begin{figure}[t]
\centerline{\includegraphics[scale=0.32]{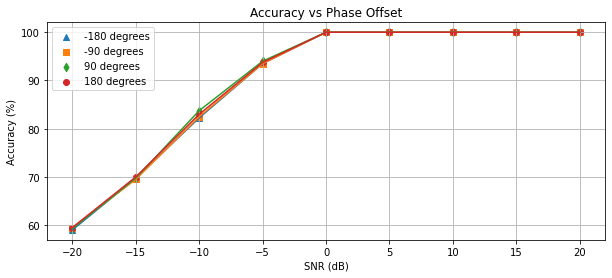}}
\caption{Impact of phase offset on classifier performance}
\label{fig5}
\end{figure}

\begin{figure}[t]
\centerline{\includegraphics[scale=0.32]{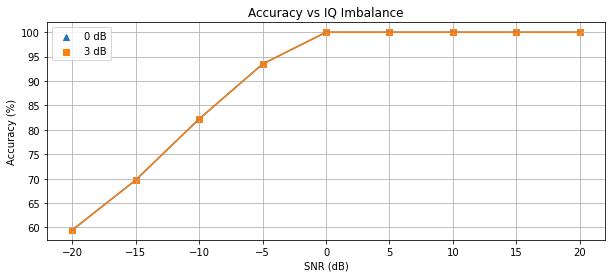}}
\caption{Impact of IQ imbalance on classifier performance}
\label{fig6}
\end{figure}

\begin{figure}[t]
\centerline{\includegraphics[scale=0.3]{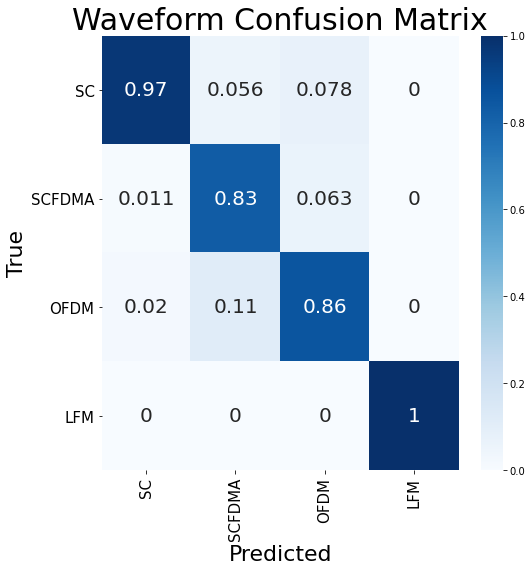}}
\caption{Confusion Matrix of all impairments from -20 dB to 20 dB}
\label{fig7}
\end{figure}

\begin{figure}[t]
\centerline{\includegraphics[scale=0.3]{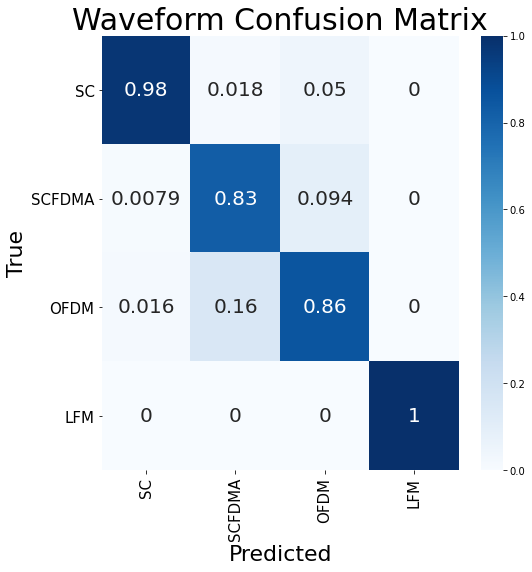}}
\caption{Confusion Matrix of AWGN impairments from -20 dB to 20 dB}
\label{fig8}
\end{figure}

\subsection{Anomaly Detector}
The results of the anomaly detector's performance are shown in Figure 9, Figure 10, and Figure 12 using only AWGN and using AWGN, multi-path fading, and other signal impairments, respectively. Three techniques were examined during testing of the detector to calculate the required RMSE thresholds to detect unknown signals. The first used a two-region design using one region for known signals and the other for unknown signals. The known region combined all known radar and communication waveforms into one class. This design used only two thresholds, found by taking the minimum and maximum of the training signal reconstruction RMSE, to reject all signals that fell outside this region. Normally, only one threshold is used to classify anything greater as unknown. However, during testing, it was found that some of the signals used for the unknown class were achieving reconstruction error lower than the maximum RMSE value, but were lower than the minimum RMSE value. To detect these signals, the minimum and maximum RMSE values were used. While the concept of the region defined by two thresholds was effective, the design use of only two regions proved to have a major disadvantage. As shown in Figure 10, there is a considerable gap between the radar reconstruction error and the communication reconstruction error. This results in half or less of the unknown signals falling in this gap and being classified as known signals during testing. Because of this, the detector performance is limited to approximately 60\% in higher SNR values as shown in Figure 9. Due to the reconstruction errors being on such a small scale, the RMSE values of the unknown signals and radar signals have to be examined without the RMSE values of the communications signals. While in Figure 10, the unknown signals appear to be in the gap. A close-up examination in Figure 11 shows the unknown signals are actually outside of the region.

The second method used a three-region design. One region is used for radar waveforms with two thresholds, one is used for communication waveforms with two thresholds to define its region as well. Signals falling in other areas besides these regions were considered to be unknown. Thresholds were found the same way as before by taking the minimum and maximum RMSE values of the radar signals and the communications signals. During testing this resulted in higher performances across the SNR range as shown in Figure 9 with greater than 95\% accuracy achieved at 0 dB. AWGN still causes loss in performance, but not to the extent seen in the two-region design.

The third design was implemented and tested to make the detector even more robust against gaps also present in the three-region design. The design divided the regions down even further into five total. This created a region for each trained waveform class. As shown in Figure 10, this method is also successful detecting unknown signals with greater than 95\% at 0 dB with the added benefit of being slightly more resilient against AWGN at low SNR values.

During the second test, it was found that the accuracy rates at 0 dB and below were not stable when using lower values of FFT sizes for the PSDs of the signals, such as 4096-FFT and 8192-FFT, once signals impaired with multi-path fading were introduced. This then led to the use of higher value sizes such as the 16384-FFT. Rates became more stable with only occasional variations in accuracy at those SNR values. While 16384-FFT still performed well with all signal impairments as shown in Figure 11, using higher FFT sizes, preferably ones closest to the minimum length of the signal samples from the four waveforms, would be more beneficial with the trade-off of added complexity. 

\begin{figure}[t]
\centerline{\includegraphics[scale=0.4]{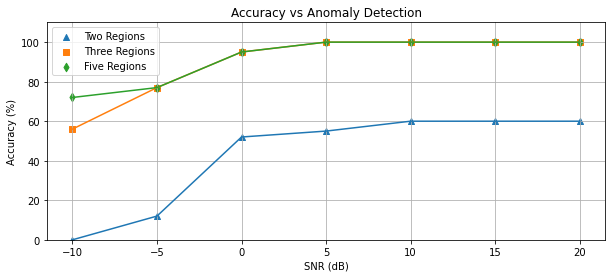}}
\caption{Overall anomaly detection performance for all three designs tested with unknown signals with AWGN using 8192-FFT}
\label{fig9}
\end{figure}

\begin{figure}[t]
\centerline{\includegraphics[scale=0.4]{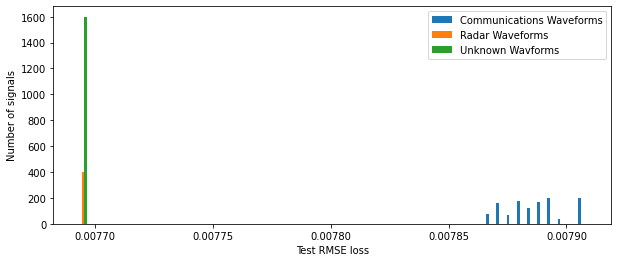}}
\caption{RMSE values of the test dataset at 20 dB using 16384-FFT}
\label{fig10}
\end{figure}

\begin{figure}[b]
    \centering
    \includegraphics[scale=0.4]{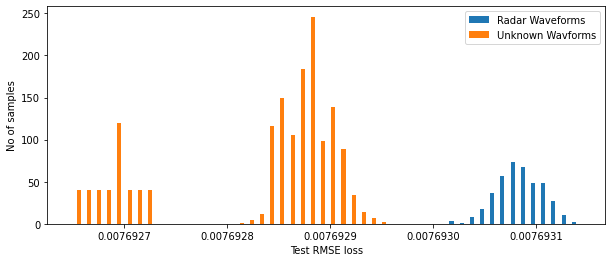}
    \caption{RMSE values of radar and unknown signals from the test dataset at 20 dB using 16384-FFT}
    \label{fig11}
\end{figure}

\begin{figure}[t]
    \centering
    \includegraphics[scale=0.4]{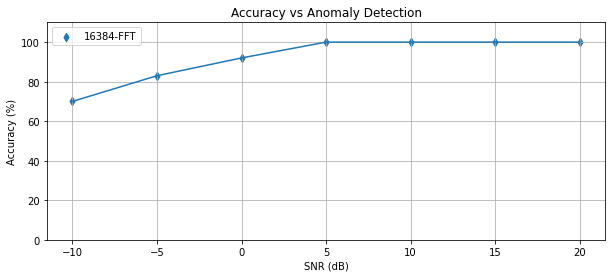}
    \caption{Performance at 16384-FFT size with fading, AWGN, and all other impairments using three-region design}
    \label{fig12}
\end{figure}

\section{Conclusion}
In this paper, we examined a method comprised of a deep feed-forward network as a waveform classifier and an autoencoder with a CNN architecture as an anomaly detector for classifying commonly used radar and communication waveforms within a shared spectrum environment. Signals are first converted to baseband, transformed to the frequency domain, and then the magnitude is taken. The anomaly detector uses the power spectral density of the signal in dB using either the same FFT size as the classifier or higher depending on desired detection sensitivity. Because of this, the conversion to frequency domain from the original signal could possibly be different for the classifier and the detector. The signal is then used as the input into the detector to determine if the signal is known. If the signal falls outside of the known regions, the system classifies the signal as unknown. If the signal is known, the signal in frequency domain before power spectral density conversion is sent to the classifier if the same FFT size is used. If not, the original signal will need to be converted to the frequency domain using the FFT size used for the classifier. The classifier then determines if the signal is a SC, SC-FDMA, OFDM, or LFM waveform.

\end{document}